%% file: sigir.tex
\documentclass[
]{ceurart}

\usepackage{listings}
\usepackage{xcolor}
\usepackage[dvipsnames]{xcolor}
\usepackage{float}
\usepackage{caption}
\usepackage{makecell}

\begin{document}

\copyrightyear{2026}
\copyrightclause{Copyright for this paper by its authors.
  Use permitted under Creative Commons License Attribution 4.0
   International (CC BY 4.0).}

\conference{ECOM'26: SIGIR Workshop on eCommerce, Jul 24, 2026, Melbourne, Australia}

\title{INSPIRE: Intent-aware Neural Sponsored Product Retrieval for E-commerce}
\author[1]{Shasvat Desai}[
email=shasvat.desai@walmart.com,
]
\cormark[1]

\author[1]{Hong Yao}
[email=Hong.Yao0@walmart.com]
\author[1]{Utkarsh Porwal}
[email=utkarsh.porwal@gmail.com]
\author[1]{Kuang-chih Lee}
[email=Kuangchih.Lee@walmart.com]
\address[1]{Walmart Global Tech, USA}

\cortext[1]{Corresponding author.}

\begin{abstract}
Walmart holds the largest share of the U.S. e-commerce grocery market, where food and beverage categories generate some of the highest search traffic and, consequently, drive a substantial portion of sponsored search revenue. At this scale, even small mismatches between user intent and retrieved products can lead to significant losses in both user engagement and monetization. Yet, understanding user intent in grocery search is inherently challenging. Queries are typically short, ambiguous, and highly diverse, often underspecifying critical preferences. For example, a query like schar white bread implicitly encodes a gluten-free preference through brand association, while queries such as chickpea pasta or oatmilk reflect underlying dietary preferences like gluten-free, plant-based, or lactose-free alternatives. Failing to capture these signals results in retrieving products that might be semantically similar but misaligned with the user’s true needs.

From the advertiser’s perspective, many products are explicitly designed to target specific intents—such as dietary preferences or size variants—and must be surfaced at the right moment to be effective. For example, a brand like Quest Nutrition, which sells high-protein, low-sugar snacks, wants its products to appear for queries like protein bars, low carb snacks, or keto snacks, even when these attributes might not be explicitly stated in the product title text. When retrieval systems fail to capture these intent signals, relevant products are not shown to the right users at the right time. From an advertiser’s perspective, this means their products are missing high-intent opportunities where conversion is most likely. Over time, this leads to lower returns on ad spend, reduced trust in the platform, and potential advertiser attrition. Losing advertisers directly translates to a loss in advertising revenue and weakens the overall sponsored search ecosystem. This challenge is further amplified in sponsored search, where only a limited number of ad slots are available, making precise relevance essential.

Thus, we propose INSPIRE (Intent-aware Neural Sponsored Product Retrieval for E-commerce), an intent-aware retrieval framework for sponsored search that leverages structured intent signals to better align user queries with relevant food and beverage products. INSPIRE represents intent as a set of structured, multi-dimensional attributes derived from both user queries and product content, capturing explicit signals (e.g., brand, flavor) as well as implicit preferences (e.g., dietary constraints, cuisine types) that are often not directly expressed in queries.
We develop a weakly supervised intent learning pipeline, where a large language model serves as a teacher to generate structured intent annotations from product titles and descriptions. We then distill these annotations by using them to finetune a lightweight student LLM model through LoRA based supervised finetuning (LoRA-SFT) that predicts intent attributes—such as brand, flavor, dietary preference, ingredient, product subtype, and cuisine type—at Walmart catalog scale. We then introduce an intent-augmented dense retrieval framework, where predicted intents are incorporated into query and product representations within a bi-encoder, enabling more precise matching between queries and sponsored products. 
To support real-world usage, we deploy the system as a scalable inference service. The distilled student model is served via a high-throughput API powered by vLLM, enabling efficient intent prediction over large product catalogs with low latency. This design ensures that intent-aware retrieval can be applied in production settings while maintaining efficiency and scalability.
\end{abstract}


\maketitle

\input{Sections/Introduction}
\input{Sections/Method}

\input{Sections/Evaluation}


\section*{Declaration on Generative AI}

During the preparation of this work, the author(s) used ChatGPT (GPT-5) and Grammarly for grammar and spelling checks. After using these tools/services, the author(s) reviewed and edited the content as needed and take full responsibility for the publication’s content.

\newpage
\bibliography{sigir}
\end{document}

%% file: Sections/Introduction.tex
\section{Introduction}
\label{sec:intro}
\begin{figure*}
    \includegraphics[width=\textwidth]
    {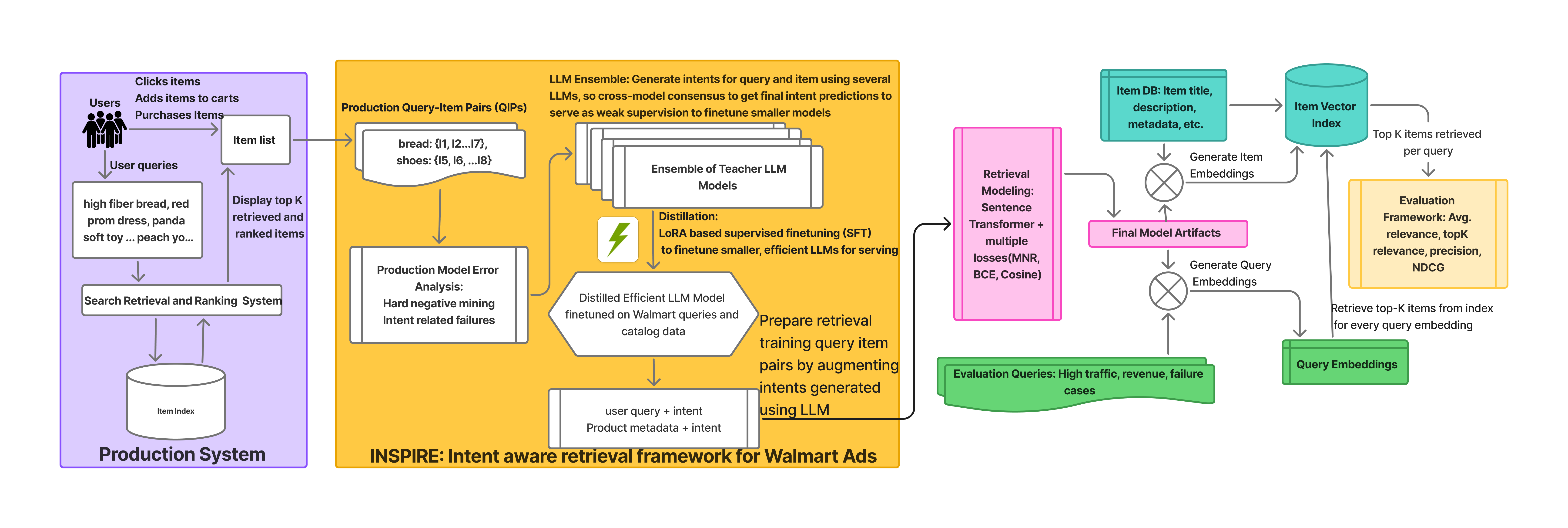}
    \vspace{-1em}
    \caption{Overview of INSPIRE, an intent-aware retrieval framework for sponsored search. The system identifies failure cases in production query–item pairs, generates structured intents using an ensemble of teacher LLMs with cross-model consensus, and distills them into efficient student models via LoRA-based supervised fine-tuning. These intents are incorporated into query and item representations to train an intent-augmented dense retrieval model, enabling improved alignment between user queries and products.}
\end{figure*}

Intent understanding for Walmart's grocery search is critical for two primary reasons: relevance and safety. Relevance is essential because Food and Beverages drives significant user traffic and revenue. Equally important is safety. Systems must respect dietary restrictions and ensure that retrieved products are allergen-safe, as failures can lead to harmful user experiences. For example, a user with a peanut allergy searching for “nut free protein bars” should not be shown products containing peanuts, even if they are otherwise highly relevant. Intent understanding requires capturing both \textit{explicit intents} (e.g., brand, flavor) and \textit{implicit intents} (e.g., dietary preferences, cuisine type). When explicit signals are ignored—such as returning a different brand for ``Coke vanilla'' or mismatching flavor in ``barbecue chips''—the results are immediately unsatisfactory and erode user trust.

The challenge is substantially greater for implicit intents, where users expect systems to infer constraints like sugar-free, dairy-free, or gluten-free even when they are only partially specified or entirely unstated. Crucially, this alignment must hold on both the \textit{query side} and the \textit{product side}: understanding what the user expresses is insufficient unless the system can also correctly interpret what the product represents. These signals are rarely explicit in product titles, requiring retrieval systems to reason over latent product knowledge rather than surface-level text. Brands often implicitly encode such properties—Schär is widely associated with certified gluten-free products, Nutpods with dairy-free and vegan offerings, and LILY’S with no-sugar-added chocolate—while ingredient-level cues further reinforce them (e.g., chickpea-based products are inherently gluten-free, Bomba rice is appropriate for paella whereas Arborio rice is not). Successfully retrieving the right items therefore demands a deeper alignment between implicit user constraints and equally implicit product attributes. As shown in Table~\ref{tab:implicit_intent_examples}, failing to capture these signals results in items that are lexically similar yet violate key constraints, making them practically unusable. Beyond preference, this also impacts safety and effort, as users are otherwise forced to inspect ingredient lists and nutritional information to verify suitability.

These challenges are further amplified in sponsored search. Ad engagement is inherently sparse—whether an ad is shown depends not only on relevance but also on campaign activity, auction dynamics, bid competitiveness, and advertiser budget. Moreover, the number of ad slots is significantly smaller than organic results, making precise relevance essential. Missing a relevant product therefore not only degrades user experience but also adversely impacts advertisers, leading to lost revenue, lower click-through rates, and reduced return on ad spend. Consequently, accurately modeling both query and product intents is critical to serving users effectively while sustaining a healthy advertiser ecosystem.

To address these challenges, we propose \textbf{INSPIRE (Intent-aware Neural Sponsored Product Retrieval for E-commerce)}, an intent-aware retrieval framework that leverages structured intent signals to better align queries with relevant food and beverage products. INSPIRE represents intent as a set of structured, multi-dimensional attributes derived from both queries and product content, capturing explicit signals (e.g., brand, flavor) as well as implicit preferences (e.g., dietary constraints, cuisine types).

We develop an INSPIRE with the following components: 
(i) \textbf{Failure identification}: perform hard negative mining and error analysis to surface query--item pairs where standard retrieval fails; 
(ii) \textbf{Intent ground truth annotation generation}: use multiple teacher LLMs to generate candidate intent ground truth labels from query and product content for weak supervision of student LLM ; 
(iii) \textbf{Consensus building}: apply cross-model agreement to filter noise and derive a high-confidence set of intent labels; 
(iv) \textbf{Model distillation}: train a smaller, efficient student LLM \cite{acharya2024survey, hinton2015distilling} via LoRA-based \cite{hu2022lora} supervised fine-tuning using high-confidence, consensus-based intent  derived from multiple teacher models as weak supervision to generate intents for Walmart catalog;
(v) \textbf{Intent aware embedding for retrieval}: incorporate these intents into the embedding generation step of the retrieval model; 
(vi) \textbf{Evaluation}: measure improvements in retrieval quality using metrics such as NDCG@K and Precision@K; 
(vii) \textbf{Deployment}: serve INSPIRE framework at scale through a vLLM-based \cite{kwon2023efficient} inference API.

\noindent
\begin{minipage}{\linewidth}
\centering
\small
\begin{tabular}{p{3cm} p{3cm} p{4cm} p{4cm}}
\toprule
\textbf{Query} & \textbf{User Intent} & \textbf{Relevant Item} & \textbf{Irrelevant Item} \\
\midrule

paella rice 
& Short-grain rice, Spanish cuisine 
& Goya Bomba Rice 
& Goya Arborio Rice \\ 
&& {\small product\_subtype: \color{OliveGreen} short grain, cuisine\_type: \color{OliveGreen} Spanish}
& {\small product\_subtype: \color{OliveGreen} short grain, cuisine\_type: \color{red} Italian} \\

\midrule

gluten free bread 
& Gluten-free 
& Schär Artisan Baker White Bread 
& Pepperidge Farm White Bread \\
&& {\small dietary\_preference: \color{OliveGreen} gluten free}
& {\small dietary\_preference: \color{red} contains gluten} \\

\midrule

sugar free chocolate 
& Low-sugar / no added sugar 
& Lily's Dark Chocolate Bar
& Hershey's Milk Chocolate Bar  \\
&& {\small dietary\_preference: \color{OliveGreen} sugar free}
& {\small dietary\_preference: \color{red} high sugar} \\

\midrule

no dairy creamer 
& Dairy-free 
& Nutpods Unsweetened French Vanilla Creamer
& Coffee Mate Original Creamer \\
&& {\small dietary\_preference: \color{OliveGreen} dairy free}
& {\small dietary\_preference: \color{red} contains dairy} \\

\midrule

gluten free protein pasta 
& Gluten-free, high protein 
& Banza Chickpea Pasta 
& Barilla Semolina Pasta
\\
&& {\small dietary\_preference: \color{OliveGreen} gluten free, protein}
& {\small dietary\_preference: \color{red} contains gluten} \\

\bottomrule
\end{tabular}
\captionof{table}{Implicit intent in grocery search is often not captured by lexical or semantic similarity alone. For example, paella specifically requires Bomba rice due to its cooking properties, while Arborio is suited for risotto. Similarly, brand-level (Schär, Nutpods, Lily’s) and ingredient-level (chickpea) knowledge encode dietary constraints such as gluten-free, dairy-free, and sugar-free. This makes it challenging to match the correct product to a query based on product names alone. Many items may appear semantically similar (red), but fail to satisfy the underlying intent, while fewer items truly align with the user’s needs (green). For instance, both Bomba rice and Arborio rice may match the query “paella rice” at a surface level, but only Bomba rice correctly satisfies the intent (green), whereas Arborio represents a misleading but similar alternative (red). Modeling these latent intent signals enables systems to distinguish between such cases and correctly rank intent-aligned products higher, while ignoring them results in items that are similar in wording but practically unusable.}
\label{tab:implicit_intent_examples}
\end{minipage}

Our contributions are as follows:
\begin{enumerate}
    \item We propose \textbf{INSPIRE}, an intent-aware retrieval framework that incorporates structured intent signals into dense retrieval to improve alignment between user queries and sponsored products. We predict the following intents as part of INSPIRE: brand, dietary preference, flavor, ingredient, cuisine type, product subtype, size value and, size unit.
    
   \item We develop a \textbf{weakly supervised labeling pipeline} that leverages large teacher LLMs and cross-model consensus to generate high-quality, structured intent annotations.

   \item We propose a \textbf{distillation framework} that transfers these intent signals into efficient student models via LoRA-based supervised fine-tuning, enabling scalable intent prediction at catalog scale.
    
    \item We demonstrate through \textbf{empirical evaluation} that intent-aware retrieval significantly improves relevance, particularly for queries involving implicit intent which cannot be effectively retrieved by semantic match.
    
    \item We have designed and are in the process of deploying a \textbf{scalable inference system} using vLLM for low-latency, high-throughput intent prediction at catalog scale.
\end{enumerate}

%% file: Sections/Method.tex

\section{Related Work}

Intent understanding has moved beyond simple taxonomies toward representation learning methods that model deeper semantic and contextual signals. Early foundations defined core navigational, informational, and transactional intents \cite{broder2002taxonomy}, followed by work leveraging behavioral signals such as click-through and session data to capture contextual intent while addressing biases in interaction logs \cite{cao2008context, chuklin2022click, joachims2017unbiased, desai2026unified}. In e-commerce, intent understanding has been studied through query modeling, term refinement, and behavioral prediction. Product-aware approaches align query representations with catalog signals, enabling the model to ground user intent in the space of available products rather than treating queries in isolation \cite{zhao2019dynamic}, while hierarchical modeling captures multi-level user needs, improving retrieval for underspecified queries \cite{ahmadvand2020jointmap}. Complementary work shows that identifying and refining intent-bearing query terms improves retrieval effectiveness by filtering noisy or non-informative tokens \cite{manchanda2019intent}, and sequential models leverage temporal user interaction patterns to predict purchase intent, particularly in session-based settings \cite{diamantaras2021predicting}. More recent approaches incorporate task-specific pre-training objectives to jointly optimize intent detection and embedding-based retrieval, leading to more semantically aligned representations \cite{qiu2022pre}.

In parallel, advances in neural retrieval have shifted from interaction-based models such as DRMM and KNRM, which rely on explicit term-level matching signals \cite{guo2016deep, xiong2017end}, to dense retrieval frameworks enabled by Transformer architectures \cite{vaswani2017attention} and Siamese encoders like Sentence-BERT \cite{reimers2019sentence}. These models map queries and documents into a shared latent space, allowing semantic matching beyond lexical overlap. Training strategies such as hard negative mining (ANCE) \cite{xiong2021ance}, large-scale dual-encoder training (DPR) \cite{karpukhin2020dpr}, and denoising with cross-encoder supervision (RocketQA) \cite{qu2021rocketqa} further improve retrieval quality, while late-interaction models like ColBERT preserve fine-grained token-level matching with efficient indexing \cite{khattab2020colbert}. More recently, large language models (LLMs) have been leveraged for deeper intent understanding and semantic alignment \cite{liu2020deep}, enabling extraction of nuanced, implicit user preferences. These models are typically adapted for production through distillation \cite{sanh2019distilbert} or parameter-efficient tuning methods such as LoRA \cite{hu2022lora}, with alignment techniques like Self-Instruct improving their ability to capture human intent \cite{wang2023self}.

However, existing approaches largely treat query intent understanding, item representation, and retrieval as loosely coupled components or rely heavily on behavioral signals. In contrast, our work jointly models structured intents for both queries and items, explicitly capturing both explicit signals (e.g., attributes, constraints) and implicit signals (e.g., brand priors, use-cases) via a multi-LLM consensus framework. We further distill these signals into an efficient student model for scalable deployment and integrate them directly into dense retrieval representations, enabling fine-grained alignment between user intent and product representations.

\section{Method}

We present \textbf{INSPIRE}, an intent-aware retrieval framework that improves alignment between user queries and sponsored products by incorporating structured intent signals derived from both query and product content. Our approach consists of five key components detailed in this section.

\subsection{Error Analysis of Query--Item Pairs from Production Logs}
We begin by analyzing production query--item pairs (QIPs) from Walmart logs, including impressions, clicks, add-to-cart, and purchase signals. We perform hard negative mining and error analysis to identify failure cases where the current retrieval system surfaces items that are semantically similar but misaligned with user intent. These failure cases highlight scenarios where implicit or explicit intent signals are not adequately captured. We analyze query- and item-level intent distributions to understand the potential impact of intent-aware retrieval.

To systematically analyze retrieval failures, we first construct a labeled set of query--item pairs (QIPs) from production logs. Each QIP is annotated with an ordinal relevance score $Rel(q,i)\in\{0,1,2,3,4\}$, corresponding to $0$ (Embarrassing), $1$ (Bad), $2$ (Okay), $3$ (Good), and $4$ (Excellent), reflecting the degree to which the item satisfies the user’s intent. These labels are obtained using a cascade of relevance models (Gemma-1B, Gemma-2B, and LLaMA-3 8B) \cite{gemma2024, touvron2023llama}, evaluated sequentially with confidence-based early exit. For uncertain hard cases where early exit is not possible, the predictions from all models are aggregated via majority voting, with ties resolved using the final-stage model. This cascade of relevance models is trained using 1.2M query-product pairs,
collected over 18 months. It was evaluated on a held-out set of 100K pairs annotated by trained raters on a
5-point scale (0=Embarrassing through 4=Excellent and class distribution approximately balanced at
18–22\% per class). The resulting relevance score is normalized to $[-1,1]$ as:
\begin{equation}
\small
\label{rel_score_eqn}
rel\_score(q,i) = (Rel(q,i)-2)/2.
\end{equation}

\paragraph{Query-side analysis.}
Approximately 60\% of queries in our evaluation set have an average relevance rating below 3, indicating that retrieved results are often only moderately relevant or poor. Among these underperforming queries, 70\% contain identifiable intent signals, and $\sim$20\% involve \textit{implicit intents} that are not explicitly expressed in the query text. Furthermore, $\sim$55\% of these queries belong to the Food category, where latent preferences such as dietary constraints and cuisine types are particularly prevalent. These findings suggest that a large fraction of retrieval failures stem from missing or misaligned intent signals, and that modeling query-side intent can significantly improve relevance.

\paragraph{Item-side analysis.}
On the product side, intent signals are highly pervasive. We observe that 98\% of food items contain at least one intent attribute. Notably, 59\% of items contain \textit{implicit intents} that are not explicitly stated in the product title, but can be inferred from brand, ingredients, or metadata. This highlights that product representations often encode rich latent signals that are not captured by surface text alone.

\paragraph{Failure case analysis.}
Table ~\ref{tab:low_performing_queries} shows examples of low-performing queries with identifiable intent signals. Many such queries contain clear intent attributes (e.g., flavor, ingredient, brand, dietary preference), yet the retrieval system fails to surface relevant items. This is especially pronounced for implicit intents (e.g., ingredient-level or cuisine-level signals), where the absence of explicit modeling leads to poor alignment despite strong lexical overlap.

These observations reveal a systematic gap: both queries and items contain rich intent signals—many of which are implicit—yet current retrieval systems fail to leverage them effectively. This motivates the need for an intent-aware retrieval framework that explicitly models and aligns query and product intents.

\subsection{Teacher LLM Intent Prediction and Consensus}
For the identified QIPs, we use an ensemble of large language models (LLMs) to generate structured intent annotations from both queries and product metadata (e.g., title, description). Each model predicts intent attributes such as brand, flavor, dietary preference, product subtype, and cuisine type. To improve robustness and reduce noise, we apply cross-model consensus, retaining only high-confidence intent predictions that are agreed upon across multiple teacher models. We construct high-quality intent annotations using a multi-stage pipeline that combines multiple LLMs, cross-model agreement, and validation.

\noindent
\begin{minipage}{\linewidth}
\centering
\small
\begin{tabular}{p{3.5cm} p{2cm} p{4.5cm} p{1.8cm} p{1.8cm}}
\toprule
\textbf{Query} & \textbf{Avg. Rel.} & \textbf{Intent (key:value)} & \textbf{Explicit} & \textbf{Implicit} \\
\midrule

soymilk vanilla & 1.086 & flavor: vanilla, ingredient: soy & flavor & ingredient \\
raspberry vinaigrette & 1.086 & flavor: raspberry & flavor & -- \\
hash brown frozen & 1.086 & storage: frozen, ingredient: potato & storage & ingredient \\
monterey jack & 1.086 & ingredient: cheese & -- & ingredient \\
bissell crosswave clean solution & 1.086 & brand: bissell & brand & -- \\
ice salt & 1.086 & ingredient: salt & -- & ingredient \\

\addlinespace

cream of mushroom soup 4 pack & 1.088 & ingredient: mushroom, package: 4 pack & package & ingredient \\
real good chicken nugget & 1.088 & ingredient: chicken & -- & ingredient \\
fruit riot & 1.088 & brand: fruit riot & brand & -- \\
hot pocket philly steak & 1.088 & flavor: philly steak & flavor & -- \\

\addlinespace

heineken light & 1.09 & brand: heineken & brand & -- \\
vegetarian sausage & 1.09 & dietary\_preference: vegetarian & dietary & -- \\
naan bread & 1.09 & cuisine\_type: indian & -- & cuisine \\
lactose free cheese & 1.09 & dietary\_preference: lactose free & dietary & -- \\

\addlinespace

vegan cream cheese & 1.092 & dietary\_preference: vegan & dietary & -- \\
pho & 1.092 & cuisine\_type: vietnamese & -- & cuisine \\
gluten free bagel & 1.092 & dietary\_preference: gluten free & dietary & -- \\

\addlinespace

black olive & 2.258 & ingredient: olive, color: black & color & ingredient \\
chicken organic & 2.258 & ingredient: chicken, dietary\_preference: organic & dietary & ingredient \\
mini m m candy & 2.258 & brand: m\&m & brand & -- \\
organic apple cider vinegar & 2.258 & dietary\_preference: organic & dietary & -- \\

\addlinespace

tcl roku television 65 & 2.26 & brand: tcl, size: 65 & brand, size & -- \\
buffalo pretzel & 2.26 & flavor: buffalo & flavor & -- \\
fresh ginger & 2.26 & ingredient: ginger & -- & ingredient \\

\bottomrule
\end{tabular}
\captionof{table}{Error analysis using log data: Examples of low-performing queries with extracted intent signals. Each query is associated with an \textit{average relevance rating across the top 25 retrieved items}, where ratings lie on an ordinal scale from $0$ (Embarrassing), $1$ (Bad), $2$ (Okay), $3$ (Good), to $4$ (Excellent). Many queries contain explicit and implicit intents, yet retrieval performance remains poor, highlighting the gap between intent presence and utilization.}
\label{tab:low_performing_queries}
\end{minipage}

\paragraph{Step 1: Multi-LLM Intent Prediction.}
For both queries and items, we obtain structured intent annotations independently from three LLMs: Gemma3 27B, LLaMA 3.1 8B, and Qwen3 8B. \cite{bai2023qwen, touvron2023llama, gemma2024}Each model produces a structured output (e.g., JSON) capturing intent attributes such as brand, flavor, ingredient, size value, size unit, product subtype, dietary preference, and cuisine type, including both explicit and implicit signals. Table ~\ref{tab:intent_gt} shows show sample intent ground truth annotations

\begin{table*}[t]
\centering
\footnotesize
\begin{tabular}{p{3.2cm} p{2.5cm} p{3.2cm} p{2cm} p{3.1cm} p{2cm}}
\toprule
\textbf{Query/Item} & \textbf{Brand} & \textbf{Dietary Preference} & \textbf{Flavor} & \textbf{Subtype} & \textbf{Size} \\
\midrule

Skinnygirl, Fat-Free, Sugar-Free Raspberry Vinaigrette Salad Dressing, 8 fl oz 
& skinnygirl
& sugar free, vegan, fat free 
& raspberry 
& raspberry vinaigrette 
& 8 fl oz \\

\addlinespace

Artisana Organics, Cashew Cacao Spread, 8 oz (227 g) 
& artisana organics
& vegan, kosher, gluten free 
& cacao 
& cashew cacao spread 
& 8 oz (227 g) \\

\addlinespace

Silk Vanilla Almond Milk 
& silk
& lactose free, vegan 
& vanilla 
& almond milk 
& -- \\

\addlinespace

Mountain Dew Zero 
& mountain dew
& sugar free 
& -- 
& soda 
& -- \\

\bottomrule
\end{tabular}
\caption{Structured attribute extraction focusing on key user-relevant intents such as brand, dietary preferences, flavor, subtype, and size.}
\label{tab:intent_gt}
\end{table*}

\paragraph{Step 2: Cross-Model Agreement and Consensus.}
To improve robustness, we compare intent annotations across models. We perform pairwise comparisons (Gemma--LLaMA, Gemma--Qwen, LLaMA--Qwen) \cite{touvron2023llama, gemma2024, bai2023qwen}and compute:
(i) token-level overlap for each intent field, and 
(ii) semantic similarity using embedding-based cosine similarity to capture lexically different but semantically equivalent predictions. 
We use predefined thresholds (0.5 for token overlap, 0.3 for embedding similarity) to determine agreement. Intent fields agreed upon by at least two models are retained as high-confidence signals, while fields with no agreement are discarded.

\paragraph{Step 3: LLM-based Verification.}
We further validate the consensus intents using a verification step with a GPT 4.1. \cite{achiam2023gpt} Given the original query/item along with the consensus and conflicting annotations, the model is prompted to validate correctness, detect hallucinations, and suggest corrections or missing implicit intents. Based on this output, intents are automatically accepted, rejected, or flagged for further review.

\paragraph{Step 4: Manual Spot-Checks.}
We perform targeted human review on low-confidence or conflicting cases to ensure annotation quality. This step helps correct edge cases and calibrate the overall pipeline.

\paragraph{Data Normalization and Synonym Mapping.}
All inputs are standardized prior to intent extraction. We apply lowercasing, whitespace normalization, and lexical normalization techniques such as stemming (e.g., ``running'' $\rightarrow$ ``run'') and lemmatization (e.g., ``better'' $\rightarrow$ ``good'', ``mice'' $\rightarrow$ ``mouse'') to reduce variability and improve consistency across model predictions.

To further ensure consistency, we normalize units by mapping surface forms (e.g., ``grams'', ``gms'', ``pounds'', ``floz'') to canonical representations (e.g., \textit{g}, \textit{lb}, \textit{oz}). This reduces sparsity and improves alignment between query and product text.

We also perform synonym normalization to capture semantically equivalent expressions that may not match lexically. Using dependency parsing, we identify modifier patterns and map them to canonical intent attributes (e.g., ``unsalted'' $\rightarrow$ \textit{salt-free}, ``unsweetened'' or ``no sugar'' $\rightarrow$ \textit{sugar-free}, ``decaf'' $\rightarrow$ \textit{caffeine-free}). 

Together, these normalization steps ensure that both explicit and implicit intent signals are represented consistently, improving the reliability of intent extraction and downstream retrieval alignment.

\subsection{Distillation via Supervised Fine-Tuning using LoRA}

We distill consensus-based intent annotations into a smaller, efficient student model (\texttt{microsoft/Phi-4-mini-instruct}) \cite{abouelenin2025phi} via supervised fine-tuning (SFT) \cite{ouyang2022training} with Low-Rank Adaptation (LoRA)\cite{hu2022lora}, enabling parameter-efficient adaptation while retaining knowledge transferred from larger teacher models. This enables scalable intent prediction while retaining knowledge encoded by larger teacher models. The resulting model predicts structured intents for both queries and products at catalog scale.

SFT trains the model on input--output pairs by minimizing the negative log-likelihood (NLL) of the target sequence conditioned on the input. Each example consists of a prompt (system + user input) and a completion (structured intent output). Inputs are tokenized and concatenated prior to training.

\noindent
\begin{minipage}{\linewidth}
\centering
\small
\begin{tabular}{p{5cm} p{4cm}}
\toprule
\textbf{Original Form} & \textbf{Canonical Form} \\
\midrule

unsalted & salt free \\
unsweetened & sugar free \\
no sugar  & sugar free \\
decaf & caffeine free \\
decaffeinated & caffeine free \\

\addlinespace

kg, kilogram, kilograms, kilo, kilos & kg \\
g, gram, grams, gm, gms, gr, grm & g \\
lb, lbs, pound, pounds & lb \\
oz, ounce, ounces, ozs, fl oz, floz, fluid ounce, fluid ounces, 0z, z & oz \\

\addlinespace

l, lt, liter, liters, ltr, litre, litres & l \\
ml, milliliter, milliliters, millilitre, millilitres & ml \\

\addlinespace

cup, cups & cup \\
pint, pints, pt, pts & pint \\
quart, quarts, qt & quart \\
gallon, gallons, gal & gallon \\

\addlinespace

packet, packets, sachet, sachets & packet \\
can, cans, tin, canister & can \\

\bottomrule
\end{tabular}
\captionof{table}{Canonical normalization of surface forms. We map diverse lexical variations—including dietary descriptors (e.g., unsweetened $\rightarrow$ sugar free) and unit expressions (e.g., grams, gms $\rightarrow$ g)—to standardized representations. This reduces sparsity and enables consistent intent extraction across queries and product descriptions.}
\label{tab:synonym_mapping}
\end{minipage}

\paragraph{Training Objective.}
We use token-level cross-entropy loss with label shifting, where the model predicts the next token given previous tokens. Padding tokens are ignored via an ignore index (set to $-100$).

We fine-tune the student model (\texttt{microsoft/Phi-4-mini-instruct}) \cite{abouelenin2025phi} using parameter-efficient Low-Rank Adaptation (LoRA) \cite{hu2022lora}. Training is performed for 3 epochs with an effective batch size of 16 (batch size 1 with gradient accumulation), using a learning rate of $4\times10^{-4}$. We use 1 H100 GPU to train this model.

We construct training examples as prompt--completion pairs, where the prompt consists of the formatted input (query or product text), and the completion corresponds to structured intent annotations. The input sequence is formed by concatenating the prompt and completion tokens.

\paragraph{Prompt Masking and Supervision.}
We apply prompt masking to ensure that the loss is computed only over the completion tokens. Specifically, tokens corresponding to the prompt are assigned an ignore index ($-100$), while completion tokens are supervised using standard cross-entropy loss. Padding tokens are also masked to avoid contributing to the loss.

Formally, given input sequence $x = [x_{\text{prompt}}, x_{\text{completion}}]$, the training objective minimizes:
\begin{equation}
\mathcal{L}_{\text{SFT}} = - \sum_{t \in \text{completion}} \log P(x_t \mid x_{<t})
\end{equation}

We tokenize prompt and completion separately and concatenate them to form a single sequence, truncated to a maximum length of 4096 tokens. We use gradient accumulation to simulate larger batch sizes under memory constraints and periodically checkpoint model weights. Padding tokens and prompt tokens are excluded from supervision to prevent degenerate learning behavior (e.g., copying prompts), ensuring that the model focuses on generating accurate intent outputs.

\subsubsection{Intent Prediction Evaluation}

We evaluate intent prediction at the attribute level using semantic matching between predicted and ground truth intents. For each intent field, we compute embeddings using a SentenceTransformer (MiniLM) model for both prediction and ground truth.

A prediction is considered a true positive if the cosine similarity exceeds a threshold of $0.6$. Based on this, we define:

\begin{itemize}
    \item \textbf{True Positive (TP):} $\text{sim}(\text{pred}, \text{GT}) > 0.6$
    \item \textbf{False Positive (FP):} prediction present but not aligned with ground truth
    \item \textbf{False Negative (FN):} ground truth present but prediction missing
    \item \textbf{True Negative (TN):} both prediction and ground truth absent
\end{itemize}

We compute precision, recall, and F1-score across all intent attributes.

\paragraph{Item Understanding Results}
The model achieves strong performance on product data, with overall precision of $0.95$ and recall of $0.97$.

\begin{center}
\small
\begin{tabular}{lcc}
\toprule
\textbf{Intent} & \textbf{Precision} & \textbf{Recall} \\
\midrule
cuisine type & 0.86 & 0.91 \\
dietary preference & 0.92 & 0.93 \\
flavor & 0.90 & 0.91 \\
brand & 0.92 & 0.90 \\
ingredient & 0.91 & 0.98 \\
product subtype & 0.98 & 0.99 \\
quantity & 0.65 & 0.64 \\
size value & 0.98 & 0.99 \\
size unit & 0.97 & 0.98 \\
\bottomrule
\end{tabular}
\end{center}

\paragraph{Query Understanding Results}
For query intent prediction, the model achieves overall precision of $0.91$ and recall of $0.93$.

\begin{center}
\small
\begin{tabular}{lcc}
\toprule
\textbf{Intent} & \textbf{Precision} & \textbf{Recall} \\
\midrule
brand & 0.98 & 0.97 \\
dietary preference & 0.94 & 0.91 \\
flavor & 0.81 & 0.91 \\
ingredient & 0.77 & 0.84 \\
size value & 0.92 & 0.89 \\
size unit & 0.93 & 0.94 \\
product subtype & 0.94 & 1.00 \\
cuisine type & 0.83 & 0.81 \\
\bottomrule
\end{tabular}
\end{center}

Next, we leverage the distilled student model to enable scalable intent-aware retrieval. While large teacher LLMs produce high-quality intent annotations, they are computationally expensive and unsuitable for large-scale inference over millions of query--item pairs. To address this, we distill their knowledge into a lightweight student model, which retains comparable intent prediction quality while enabling low-latency, high-throughput inference.

\subsection{Retrieval Data Augmentation}
\label{sec:intent_augmentation}
We use the distilled student model to predict structured intents for both queries and products at scale. These predicted intents are then used to augment the retrieval training data. 

\noindent
\begin{minipage}{\linewidth}
\centering
\small
\begin{tabular}{p{3.0cm} p{4.5cm} p{4.5cm} p{4.5cm}}
\toprule
\textbf{Representation} & \textbf{Query} & \textbf{Relevant Item} & \textbf{Irrelevant Item} \\
\midrule

\textbf{Original} 
& gluten free bread 
& Schär Artisan Baker White Bread 
& Pepperidge Farm White Bread \\

\textbf{Intent-Augmented} 
& gluten free bread ; {\small dietary preference: gluten free}
& Schär Artisan Baker White Bread ; {\small dietary preference: gluten free, product subtype: bread}
& Pepperidge Farm White Bread ; {\small dietary\_preference: contains gluten, product\_subtype: bread} \\

\addlinespace
\midrule

\textbf{Original} 
& no dairy creamer
& Nutpods French Vanilla Creamer 
& Coffee Mate Original Creamer \\

\textbf{Intent-Augmented} 
& no dairy creamer ; {\small dietary preference: dairy free}
& Nutpods French Vanilla Creamer ; {\small dietary preference: dairy free, flavor: vanilla}
& Coffee Mate Original Creamer ; {\small dietary preference: contains dairy} \\

\addlinespace
\midrule

\textbf{Original} 
& paella rice
& Goya Bomba Rice 
& Goya Arborio Rice \\

\textbf{Intent-Augmented} 
& paella rice ; {\small cuisine type: Spanish, product subtype: short grain rice}
& Goya Bomba Rice  ; {\small cuisine type: Spanish, product subtype: short grain rice}
& Goya Arborio Rice ; {\small cuisine type: Italian, product subtype: short grain rice} \\

\bottomrule
\end{tabular}
\captionof{table}{Intent-aware data augmentation. Adding structured intent attributes to both queries and products enables explicit alignment of user constraints with product characteristics, improving discrimination between relevant and lexically similar but unsuitable items. For example, for the query \textit{gluten free bread}, it is not immediately clear which product is a better match just based on the title:  \textit{Schär Artisan Baker bread} or \textit{Pepperidge bread}. However, once we augment the intent key values, it becomes more obvious -  the relevant item shares the attribute \textit{{\small dietary preference: gluten free}}, whereas the irrelevant item contains gluten despite similar surface form. This differentiaion enables higher lexical and semantic match between query and relevant item, enabling us to improve relevance of retrieved items}
\label{tab:intent_augmentation}
\end{minipage}

\subsection{Retrieval Model Training}

We train an intent-aware dense retrieval model that leverages structured intent signals to better align user queries with relevant products. Our approach consists of three key components: (i) constructing a unified supervision signal for query--item pairs that captures both semantic relevance and user preference, (ii) augmenting queries and items with predicted intent attributes to enable explicit intent-level matching, and (iii) training a bi-encoder retrieval model using these enriched representations and supervision signals. We describe each of these components in detail below. 

\paragraph{QIP Supervision Signal.}
We train the retrieval model on query--item pairs (QIPs) $(q,i)$ constructed from production logs, capturing real user interactions. Each QIP is assigned a continuous supervision score that reflects both semantic relevance and user preference. We train the retrieval model on query--item pairs (QIPs) $(q,i)$, each assigned a continuous supervision score capturing both semantic relevance and user preference. 

Each QIP is first assigned an ordinal relevance label $Rel(q,i)\in\{0,1,2,3,4\}$, representing the degree of intent fulfillment from \textit{Embarrassing} to \textit{Excellent}. These labels are obtained via a cascade of cross-encoder teacher models (Gemma-1B, Gemma-2B, LLaMA-3 8B) along with available human annotations. We normalize this label to:
\begin{equation}
rel\_score(q,i) = \frac{Rel(q,i) - 2}{2} \in [-1,1]
\end{equation}

We further incorporate user engagement signals derived from aggregated interactions (orders, add-to-cart, clicks, views). These signals are log-compressed, normalized per query, and smoothed to produce $\tilde{E}(q,i)$. Engagement is incorporated only for semantically relevant QIPs ($Rel(q,i) \geq 2$) to avoid promoting irrelevant but popular items.

The final supervision score is defined as:
\begin{equation}
y(q,i) =
\mathrm{clip}\left(\mu_{\text{rel}} \cdot rel\_score(q,i) + \lambda_{\text{eng}} \cdot \tilde{E}(q,i), 0, 1\right)
\end{equation}

This formulation ensures that supervision reflects both intent-level semantic alignment and real user preferences.

As described in Section~\ref{sec:intent_augmentation}, to enable intent-aware retrieval, we augment both queries and items with structured intent attributes predicted by the distilled student model. Let $x_q$ and $x_i$ denote the original query and item text, and $\mathcal{I}(q)$ and $\mathcal{I}(i)$ denote their predicted intent sets. We construct:
\begin{equation}
\tilde{x}_q = x_q \;\Vert\; \mathcal{I}(q), \quad
\tilde{x}_i = x_i \;\Vert\; \mathcal{I}(i)
\end{equation}

This augmentation enables the model to explicitly capture alignment between user constraints (e.g., dietary preferences, cuisine type) and product attributes.

\paragraph{Training Data.}
Our training dataset consists of approximately $10$M QIPs derived from production logs, combining relevance-labeled pairs and engagement-based signals. This large-scale dataset provides diverse supervision across both explicit and implicit intent scenarios.

\paragraph{Model and Training Objective.}
We adopt a bi-encoder architecture based on a MiniLM SentenceTransformer. \cite{reimers2019sentence, wang2020minilm, wolf2019huggingface} Queries and items are encoded independently:
\begin{equation}
\mathbf{h}_q = f(\tilde{x}_q), \quad \mathbf{h}_i = f(\tilde{x}_i)
\end{equation}

Relevance is computed via cosine similarity:
\begin{equation}
s(q,i) = \cos(\mathbf{h}_q, \mathbf{h}_i)
\end{equation}

We train the model using a combination of Multiple Negatives Ranking (MNR) loss \cite{xiong2021ance} and cosine regression loss:
\begin{equation}
\mathcal{L} = \mathcal{L}_{\text{MNR}} + \lambda \cdot \mathcal{L}_{\text{cosine}}
\end{equation}

The MNR loss enforces separation between positive and in-batch negative pairs, while the cosine loss aligns predicted similarity scores with the continuous supervision target $y(q,i)$. This enables the model to distinguish between items that are lexically similar but differ in intent-level compatibility.

%% file: Sections/Evaluation.tex
\section{Evaluation}
We evaluate the effectiveness of INSPIRE using standard retrieval metrics, including NDCG@K, Precision@K, and average relevance@K. Evaluation is performed on a curated set of high-traffic queries, revenue-driving queries, and known failure cases, demonstrating improvements in alignment, particularly for queries involving implicit intent. We are still in progress of deploying this system and plan to run an A/B test after the deployment.
\subsection{Evaluation Protocol}
\label{sec:eval_protocol}

We construct an evaluation set of 12000 queries that covers both head and tail traffic regimes. Queries are sampled from (i) high-traffic segment, (ii) high-revenue segments, (iii) long-tail queries, and (iv) historically under-performing queries. We build an item index over the full item inventory using FAISS \cite{johnson2019billion} and retrieve the top-$25$ candidates for each query. We report metrics at $K{=}25$ because the sponsored search page has a fixed number of ad slots; in our setting, $25$ is the optimal number of ad items for our application. Each retrieved query--item pair is annotated on a 5-point relevance scale. We use a hybrid judging setup consisting of (i) available human annotations and (ii) model-based annotations from a strong DeBERTa cross-encoder relevance model fine tuned on internal data for pairs without human labels. Out of the 300,000 query-item pairs (QIPs) that we evaluate the model on, human annotations were obtained for 100K QIPs.
We use the resulting 5-point labels to compute graded retrieval metrics. We report (i) average relevance score over the retrieved list, (ii) Precision@$K$, and (iii) NDCG@$K$

\subsection{Quantitative Evaluation}
Intent-aware retrieval consistently improves ranking quality across all evaluation metrics. As shown in Table~\ref{tab:intent_retrieval_results}, average relevance@10 increases from 3.033 to 3.105 (+2.38\%), and average relevance@25 improves from 3.01 to 3.08 (+2.3\%), indicating better overall alignment between retrieved items and user intent, both at top ranks and deeper in the list.

\begin{table}[t]
\centering
\small
\setlength{\tabcolsep}{4pt}
\renewcommand{\arraystretch}{1.1}
\begin{tabular}{l c c c}
\toprule
\textbf{Metric} & \textbf{Non-Intent} & \textbf{Intent-Aware} & \textbf{Relative Gain} \\
\midrule

Avg relevance@10 & 3.033 & 3.105 & \color{OliveGreen} +2.38\% \\
Median relevance@10 & 3.0 & 3.0 & -- \\

Avg relevance@25 & 3.01 & 3.08 & \color{OliveGreen} +2.3\% \\
Median relevance@25 & 3.0 & 3.0 & -- \\

\addlinespace
\multicolumn{4}{l}{\textit{Precision@K (Rel $\geq$ 3)}} \\
\midrule
Precision@1  & 0.812 & 0.846 & \color{OliveGreen} +4.2\% \\
Precision@10 & 0.767 & 0.798 & \color{OliveGreen} +4.0\% \\
Precision@25 & 0.741 & 0.768 & \color{OliveGreen} +3.6\% \\

\addlinespace
\multicolumn{4}{l}{\textit{NDCG@K}} \\
\midrule
NDCG@1  & 0.903 & 0.932 & \color{OliveGreen} +3.2\% \\
NDCG@10 & 0.872 & 0.895 & \color{OliveGreen} +2.64\% \\
NDCG@25 & 0.854 & 0.874 & \color{OliveGreen} +2.3\% \\

\addlinespace
\multicolumn{4}{l}{\textit{Relevance Distribution (\%)}} \\
\midrule
Score = 4 (Excellent) & 43.0 & 47.3 & \color{OliveGreen} +10.0\% \\
Score = 3 (Good)      & 32.6 & 33.7 & \color{OliveGreen} +3.4\% \\
Score = 2 (Okay)      & 8.4  & 5.5  & \color{OliveGreen} -34.5\% \\
Score = 1 (Bad)       & 13.0 & 12.0 & \color{OliveGreen} -7.7\% \\
Score = 0 (Embarrassing) & 3.0 & 1.5 & \color{OliveGreen} -50.0\% \\

\bottomrule
\end{tabular}
\caption{Offline retrieval performance comparison between baseline (Non-Intent) and intent-aware retrieval. Intent-aware modeling improves ranking quality across all metrics, with the largest gains at top ranks and a clear shift toward higher-quality results.}
\label{tab:intent_retrieval_results}
\end{table}

We also observe consistent gains in standard IR metrics. Precision@K improves across all cutoffs, with the largest gains at top ranks (Precision@1: +4.2\%), indicating that intent-aware modeling is particularly effective at improving early precision. Similarly, NDCG@K improves across all ranks (NDCG@10: +2.64\%, NDCG@25: +2.3\%), demonstrating that relevant items are not only retrieved more often but are also ranked higher.

Importantly, intent modeling shifts the relevance distribution toward higher-quality results. The proportion of \textit{Excellent} items (score 4) increases by +10.0\%, while \textit{Good} items (score 3) also see a modest increase (+3.4\%). In contrast, mid- and low-quality results are significantly reduced: score 2 (Okay) decreases by -34.5\%, score 1 (Bad) by -7.7\%, and score 0 (Embarrassing) by -50.0\%. This indicates that incorporating structured intents helps eliminate semantically similar but constraint-violating items

Overall, these results validate that intent-aware representations lead to more precise, better-ranked, and more user-aligned retrieval outcomes.

\subsection{Qualitative Evaluation}
\begin{table*}[t]
\centering
\small
\setlength{\tabcolsep}{3pt}
\renewcommand{\arraystretch}{1.2}
\begin{tabular}{p{2cm} p{2.5cm} p{4.2cm} c p{4.2cm} c}
\toprule
\textbf{Setting} & \textbf{Query} & \textbf{Irrelevant Item} & \textbf{Score} & \textbf{Relevant Item} & \textbf{Score} \\
\midrule

Without Intent 
& \textit{peanut free snack} 
& Planters Roasted Peanuts 
& 0.57 
& MadeGood Granola Bars 
& \color{Red} 0.38 \\

With Intent 
& \textit{peanut free snack} 
& Planters Roasted Peanuts 
; {\small dietary\_preference: contains peanuts}
& 0.49
& MadeGood Granola Bars 
; {\small dietary\_preference: peanut free}
& \color{OliveGreen} 0.58 \\

\addlinespace
\midrule

Without Intent 
& \textit{gluten free pasta} 
& Barilla Farfalle Pasta 
& 0.56 
& Banza Chickpea Pasta 
& \color{Red} 0.37 \\

With Intent 
& \textit{gluten free pasta} 
& Barilla Farfalle Pasta ; {\small dietary\_preference: contains gluten}
& 0.51
& Banza Chickpea Pasta 
; {\small dietary\_preference: gluten free}
& \color{OliveGreen} 0.88 \\

\addlinespace
\midrule

Without Intent 
& \textit{soy free hamburger buns} 
& Nature's Own Brioche Buns 
& 0.50 
& Canyon Bakehouse Buns 
& \color{Red} 0.55 \\

With Intent 
& \textit{soy free hamburger buns} 
& Nature's Own Brioche Buns 
; {\small dietary\_preference: contains soy}
& 0.57 
& Canyon Bakehouse Buns 
; {\small dietary\_preference: soy free}
& \color{OliveGreen} 0.89 \\

\addlinespace
\midrule

Without Intent 
& \textit{sugar free chocolate} 
& Hershey's Milk Chocolate 
& 0.61 

& Lily's Dark Chocolate 
& \color{Red} 0.44 \\

With Intent 
& \textit{sugar free chocolate} 
& Hershey's Milk Chocolate 
; {\small dietary\_preference: contains sugar}
& 0.45
& Lily's Dark Chocolate 
; {\small dietary\_preference: sugar free}
& \color{OliveGreen} 0.84 \\

\addlinespace
\midrule

Without Intent 
& \textit{no dairy creamer} 
& Coffee Mate Original 
& 0.59 
& Nutpods Creamer 
& \color{Red} 0.46 \\

With Intent 
& \textit{no dairy creamer} 
& Coffee Mate Original 
; {\small dietary\_preference: dairy}
& 0.51
& Nutpods Creamer 
; {\small dietary\_preference: dairy free}
& \color{OliveGreen} 0.86 \\

\bottomrule
\end{tabular}
\caption{Impact of intent augmentation on retrieval scoring. Each query is shown under two settings: without intent (top) and with intent (bottom). In the absence of intent signals, the model relies primarily on lexical and semantic similarity, often assigning higher scores to semantically similar but constraint-violating items and lower score to relevant items(red). After augmenting item representations with structured intent attributes (e.g., dietary\_preference), the model prioritizes items that satisfy both explicit and implicit user constraints by scoring them higher than the irrelevant one. Relevant items that were previously under-ranked receive higher scores after intent augmentation (green), demonstrating improved alignment between user intent and retrieved products.}
\label{tab:qualitative_intent_results}
\end{table*}
Table~\ref{tab:qualitative_intent_results} provides qualitative examples illustrating how intent-aware modeling corrects common retrieval failures. 

Without intent modeling, the system often favors lexically or semantically similar items that violate user constraints. For instance, for the query \textit{``peanut free snack''}, a peanut-based product is incorrectly scored higher due to lexical overlap. Similarly, for \textit{``gluten free pasta''}, traditional wheat pasta is preferred over a gluten-free alternative.

With intent augmentation, the model explicitly captures structured constraints such as \textit{dietary\_preference} (e.g., peanut-free, gluten-free, soy-free). This enables correct ranking of items that satisfy user requirements, such as allergen-free snacks, and chickpea-based pasta

These examples highlight that intent-aware retrieval enables constraint-level reasoning beyond surface similarity, leading to more accurate and user-aligned results.

Together, the quantitative and qualitative results demonstrate that modeling structured intents bridges the gap between intent presence and intent utilization in retrieval systems.

\section{Serving Architecture}

We are in process of developing and deploying this system to production and plan to run A/B test after that. At serving time, our system combines pre-computed item and query intents to enable intent-aware retrieval.

\subsection{Offline Intent Generation Pipelines}

\begin{figure*}[t]
    \centering
    \includegraphics[width=\textwidth]{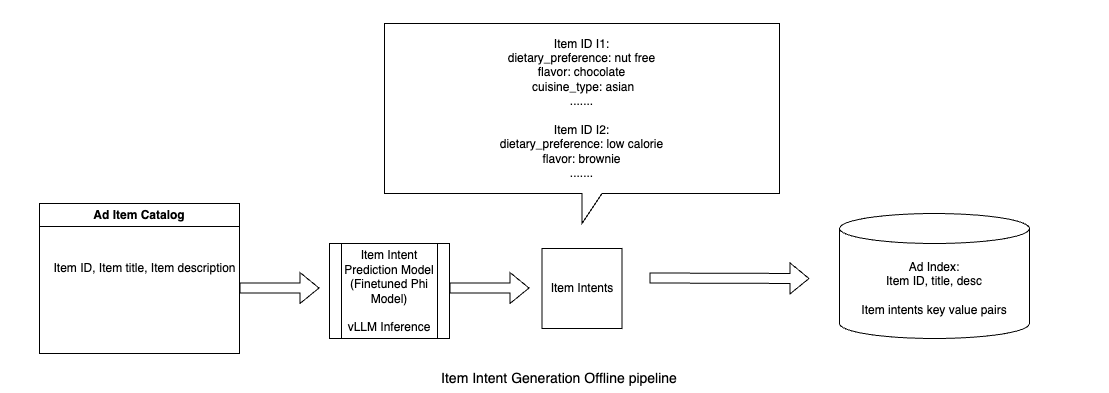}
    \caption{Offline pipelines for generating structured intents for items and queries. Item intents are extracted from catalog metadata and stored in the index. Query Intents come from an upstream intent prediction model(not shown here).}
    \label{fig:intent_pipeline}
\end{figure*}

As shown in Figure ~\ref{fig:intent_pipeline}, we run a large-scale vLLM-based inference pipeline over the entire catalog to generate structured \textit{item intents} (e.g., dietary preference, flavor, cuisine). We plan to store these intents in an \textit{Item Intent Store} and inject them into the item index as additional structured features. On the query side, we  plan to leverage intents predicted by another team to avoid duplication of effort.  


\paragraph{Item intent generation.}
Each item (title and description) from the catalog is processed using a fine-tuned LLM to extract structured attributes such as dietary preferences, flavor, and cuisine type. These intents are stored as key-value pairs and incorporated into the item index. To keep the index up-to-date, we run incremental refreshes on newly added or updated items every hour, and a full refresh over the entire catalog on a weekly basis. This step is performed offline to ensure scalability and avoid serving-time overhead.



\subsection{Online Pipeline}

\begin{figure*}[t]
    \centering
    \includegraphics[width=\textwidth]{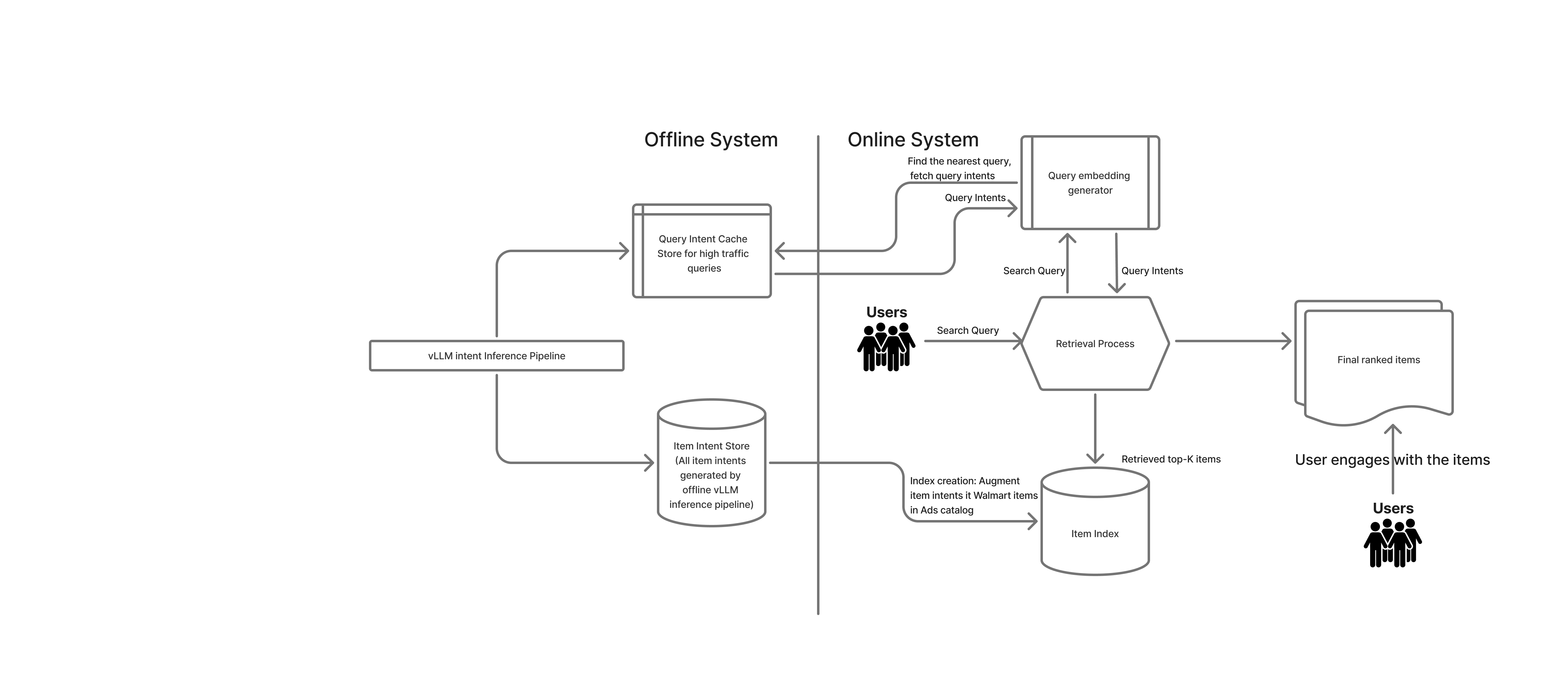}
    \caption{End-to-end serving architecture for intent-aware retrieval. The offline system generates item intents and caches query intents for frequent queries, while the online system performs real-time query intent inference and retrieves candidates from an intent-augmented index to produce final ranked results.}
    \label{fig:serving_architecture}
\end{figure*}

As shown in Figure~\ref{fig:serving_architecture}, when a user issues a query, we check the if the normalized query exists(using
embedding lookup) in the Query Intent Cache. As part of phase-1, we plan to use exact match(ENN) with the possibility to use approximate nearest neighbor match(ANN) to expand the query converage. The query, along with its inferred intents, is passed into the retrieval system, which leverages both semantic signals and intent alignment.

\section{Limitations and Future Work}

While our approach demonstrates consistent improvements in retrieval quality, it has several limitations. First, the paper only provides offline results. We plan to run A/B test after deploying this system to production and run an online test.  Second, the quality of intent prediction is bounded by the teacher models used in the distillation pipeline. Errors or biases in the LLM-generated annotations can propagate to the student model, particularly for ambiguous or long-tail queries. Third, our framework relies on a predefined schema of structured intents, which may not fully capture the richness and evolving nature of user needs, especially in open-ended or cross-domain queries. Fourth, although we distill the model for scalability, maintaining up-to-date intent annotations for rapidly changing catalogs introduces additional system overhead. 

In terms of future directions, one promising avenue is to move beyond fixed schemas toward more flexible or dynamic intent representations that can adapt to new domains and emerging user behaviors. Incorporating user context—such as session signals or personalization—can further improve intent disambiguation, particularly for underspecified queries. Another direction is to explore tighter integration between intent modeling and ranking, for example through joint training objectives or end-to-end optimization. Finally, improving the robustness of intent extraction through better alignment techniques, multi-model consensus strategies, or human-in-the-loop validation could further enhance reliability and downstream performance.